\documentclass[12pt]{article}
\usepackage{amsmath,amssymb,cite}

\newcommand{\ad}{{\rm adj}}
\newcommand{\adj}{{\bf adj\,}}
\renewcommand{\c}{{\rm c}}
\newcommand{\ch}{{\rm ch}}
\newcommand{\F}{{\mathbb F}}
\renewcommand{\P}{{\mathbb P}}
\newcommand{\Z}{{\mathbb Z}}

\newcommand{\ord}{{\rm ord\,}}
\renewcommand{\O}{{\cal O}}

\renewcommand{\L}{{\cal L}}

\newcommand{\tr}{{\rm tr}}
\newcommand{\V}{{\cal V}}

\begin{document}
\title{
\vbox{
\baselineskip 14pt
\hfill \hbox{\normalsize
} \\
\hfill \hbox{\normalsize KUNS-2234} } \vskip 2cm Extended Gauge
Symmetries in F-theory}
\author{Kang-Sin Choi\footnote{email:
  kschoi@gauge.scphys.kyoto-u.ac.jp} \\
{\it \normalsize
Department of Physics, Kyoto University,
Kyoto 606-8502, Japan}
}
\date{}
\maketitle

\thispagestyle{empty}
\begin{abstract}
We study gauge symmetry in F-theory in light of global aspects.
For this, we consider not only a
simple (local) group, but also a semi-simple group with Abelian
factors. Once
we specify the complete gauge group by decomposing the discriminant,
analogous to arranging 7-branes, we can derive the matter contents,
their localization and the relation to enhanced groups. Global
constraints coming from Calabi--Yau conditions and anomaly
cancellations imply a unified group.
The semisiple group shows explicit formation of matter curves
and nontriviality of its embedding into exceptional group. 
Also the dual heterotic string vacua with line bundles provide a guide
on the unification.
\end{abstract}

\newpage

\section{Introduction}

We study gauge symmetry in F-theory focusing on the global nature of
7-branes. F-theory \cite{Vafa:1996xn,MV1,MV2} is a good unified
framework where gauge symmetry is readily described by branes
generalized from those of open string, as well as heterotic dual is
directly established. Also exceptional group is naturally obtainable
in both pictures. On the F-theory side, we have accessible tool for
describing gauge group in terms of singularity of compact space of
the same name, sharing the same connectedness. In the perturbative
limit, the singularity is interpreted as $(p,q)$-branes and
orientifold planes \cite{Sen,GP}. The matter contents are obtained
by purely geometric way, e.g. from the intersections of 7-branes. On
the heterotic side, a similar brany description is translated to the
information on instantons \cite{Bershadsky:1997zs,FMW}. On both
sides, the moduli space and the spectrum are independently
well-understood \cite{Beetal,KV,Duff:1996rs}.

So far, much attention is paid on a unified model based on a simple
group in the local picture. The decoupling limit $M_{\rm Pl}/M_{\rm
GUT} \to \infty$ allows us to concentrate on one local group and to
make easy bottom-up construction
\cite{BHV,DW,Hayashi:2008ba,Bourjaily:2009vf}. On the other hand, it
is also noted that the matter contents and their interactions are
accounted by gauge symmetry enhancement. This is naturally explained
by further unification to a larger group. The question, to what
extent and what kind of enhancement is possible, leads us to seek
the global structure.

On the heterotic side, a simple gauge group occurs under the
instanton background as a nonabelian gauge bundle. At the generic
point of moduli space, this group can be the only surviving one,
hence in the sense of heterotic duality, the local picture with a
single simple group is 
also globally consistent. However, this is not the only possibility.
In the low energy limit of F-theory, we have an adjoint scalar
$\phi_{mn}$, spatially transforming as the canonical divisor on the
compact space, parameterizing the normal direction to the 7-branes
\cite{BHV}, which is similar to the scalar for the D7-brane in the
perturbative picture. A nonzero vacuum expectation value (VEV) of
this field gives rise to brane separation, and a non-constant
expectation value makes the branes intersecting. As a result, we can
have more than one simple or Abelian subgroup. Again, on the
heterotic side, a line bundle background that gives rise to
semisimple and Abelian gauge group is crucial to understand. Because
the line bundle commutes with every other group, the rank is not
reduced, giving rise to more fruitful spectrum
\cite{Nibbelink:2007rd,Blumenhagen:2005pm}. For this, in Section 2,
we extend the work of Ref. \cite{Beetal} to access the full
decomposition of discriminant, using intersection theory. It is
powerful for describing more than one group on the equal footing,
contrary to the Weierstrass form that can see only one group at once
in general. 

We will see that there are constraints from the Calabi--Yau
condition and anomaly cancellations, so that we cannot have
arbitrary large group. Usually this fact indicates that we can
understand at least a certain class of vacua as broken symmetry of
some unified group \cite{Un}. In Section 2, we also analyze the
spectrum and moduli space and seek relations to some unification
group. Also we can think about the opposite direction of obtaining a
realistic model from a given unified group.

Furthermore, there are some objects and observables that are to be
{\em derived} from top-down information
\cite{AG,HKTW,Donagi:2009ra,Maetal}. In the local picture, the
matter curve, where a matter of a certain representation is
localized, is assumed to wrap on a certain subcycle on the brane. If
we specify a Calabi--Yau manifold, the background on this manifold
should give us sufficient information on low-energy spectrum and
coupling, calculated by index theorem \cite{BHV,DW}. It is done by
relating geometric data to Calabi--Yau information and specifying
the cycles supporting the gauge group, constrained by global
consistency condition. If we then know how the gauge symmetry is
broken, in principle we should be able to calculate the real
location of the matter curves. This is the case in the perturbative
picture, we have the clear answer that a chiral matter under the
bifundamental representation is localized at the intersections of
two stacks of D-branes. Generalization of such mechanism to
exceptional groups is known in mathematics literatures \cite{KM,KV},
however its application to physics have been limited
\cite{Bershadsky:1996nu,Louis:1996mt,Candelas:1997eh}. We look for
such top-down picture in the F-theory description, where a
nontriviality comes from the exceptional gauge group. And Yukawa
coupling involves the normalization of matter wavefunctions along
the entire volume of the cycle they live on
\cite{HKTW,AC,Font:2009gq}. In Section 2 we consider the gauge
sector, mainly in six dimension where the only necessary global
consistency condition is 7-brane charge conservation
\cite{MV1,MV2,Sa}. In Section 3, we study such derived objects from
the 7-branes. Many examples, especially semisimple groups, are dealt
with in Section 4. We conclude in Section 5.

\section{Gauge fields on 7-branes}

\subsection{Description}
At present, the only possible definition of F-theory is via type IIB
string theory. Namely, we identify the $SL(2,\Z)$ symmetric
axion-dilaton field $\tau$ of IIB with the complex structure of an extra
torus, lifting the theory twelve-dimensional
\cite{Vafa:1996xn}.\footnote{See Refs. \cite{Johnson:2000ch,K3} for general introduction.} This
requires $X_4$ to be an elliptic fibration over a
three-dimensional base $B_3'$, $\pi': X_4 \to B_3'$.
With a section, the elliptic fiber admits description in terms of
Weierstrass equation
\begin{equation} \label{Weieq}
 y^2 = x^3 + f x + g,
\end{equation}
where $f$ and $g$ are polynomials on $B_3'$.
For the total space $X_4$ to be Calabi--Yau, $f$ and $g$ transform as
holomorphic sections of $-4 K_{B_3'}$ and $-6 K_{B_3'}$ respectively, where
$K_{B_3'}$ is the canonical class of the base manifold
  $B_3'$ \cite{Naka}.\footnote{We use the
same name for a divisor, its associated line bundle and
the first Chern class of this line bundle.}  Roughly it means $f$ and
$g$ can be locally viewed as function of coordinates in
$B_3'$,
globalizing with topological numbers 4 and 6, respectively.
Due to the $SU(4)$ special holonomy of $X_4$, the compacfication
leaves 1/8 of the supersymmetry, which is ${\cal N} = 1$ in terms of
four dimensional supersymmetry.
The complex structure $\tau$ of the elliptic (torus) fiber is related to $f,g$
through a modular function called the $j$-function \cite{Vafa:1996xn},
\begin{equation} \label{Ddef}
 j(\tau) = {4(24 f)^3 \over \Delta}, \quad \Delta \equiv 4 f^3 + 27 g^2.
\end{equation}
The discriminant locus $D$ is a divisor, or collection of codimension
one subspaces,
of $B_3'$ specified by the equation $\Delta =0$.
$\Delta$ transforms as a section of $-12 K_{B_3'} = 12 \c_1(B_3')$
and is expressed as a formal sum of
irreducible divisors of $B_3'$
\begin{equation} \label{Ddecomp}
 D = -12 K_{B_3'} = \sum_i n_i S_i, \quad n_i \ge 0.
\end{equation}
Later, we will check to what extent this form is meaningful.

\begin{table}[t]
$$\begin{array}{|c|c|c|c|c|}
\hline
\ord f & \ord g & \ord \Delta &\mbox{fiber}&\mbox{group}\\
\hline
\geq0&\geq0&0&{\rm I}_0& \cdot \\
0&0&n&{\rm I}_n&A_{n-1}\\
\geq 1&1&2&{\rm II}& \cdot \\
1&\geq2&3&{\rm III}&A_1\\
\geq 2&2&4&{\rm IV}&A_2\\
 2&\geq 3&n+6&{\rm I}^*_n&D_{n+4}\\
\geq 2& 3& n+6&{\rm I}^*_n&D_{n+4}\\
\geq 3&4&8&{\rm IV}^*&E_6\\
3&\geq 5&9&{\rm III}^*&E_7\\
\geq 4&5&10&{\rm II}^*&E_8\\
\hline
\end{array}$$
\caption{Kodaira classification of singularities.}
\label{t:fibclass}
\end{table}

Going close to {\em one} $S_i$, the elliptic fiber degenerates and gives
rise to a singularity in the torus, in the sense of (\ref{Ddef})
with $\Delta \to 0$. The corresponding curve at $S_i$ locally looks
like $y^2 = x^3 + f_i x + g_i$ and carries the orders
$\ord(f_i,g_i,\Delta_i)$, displayed in Table \ref{t:fibclass}. In
the relation (\ref{Ddecomp}), $n_i = \ord \Delta_i$  at the
corresponding $S_i$ \cite{MV1}. It gives rise to the world-volume
gauge group of the same name. In string theory language, each $S_i$
provides four-cycle which a number of 7-branes wrap, which carries
some units of RR and NSNS charges. In case of D-branes, only the
fundamental open string with two ends can end on D-branes, so that
the possible gauge groups are of $U,SO,Sp$ types. However if a
7-brane carries NSNS charge as well, there can be zero modes of a
tensionaless string junction with more than two endpoints
\cite{Junction,Choi:2005pk}, so the gauge symmetry can be enhanced
to exceptional group.

It is useful to understand this in terms of intersection
theory. Indentifying the gauge group on $S_i$ by $\ord(f,g,\Delta)$,
we write
\begin{equation} \label{matbr} \begin{matrix}
 F &=& -4 K_{B'} &=& (\ord f_i) S_i &+& F', \\
 G &=& -6 K_{B'} &=& (\ord g_i) S_i &+& G' ,  \\
 D &=&  -12 K_{B'} &=& (\ord \Delta_i) S_i&+&
  \displaystyle \sum_R s_{R_i} D'_{R_i}.
\end{matrix}
\end{equation}
This decomposition $F,G,D$ respectively shows the information on {\em
  dominant} terms in $f,g,\Delta$.
The last term in $D$ is the consequence of Tate's algorithm
\cite{Beetal,Ta}. Each matter $R_i$ is localized on $S_i \cdot
D'_{R_i}$, which is directly interpreted as the multiplicity of
$R_i$ in six dimensions. In geometric engineering, a gauge group is
not only determined by a singularity type but also a monodromy
condition \cite{AG}. The redundancy factors $s_{R_i}$, which we call
splitness, contain the information, some of which are also displayed
in Table \ref{t:invs}. Later we will reconstruct $s_{R_i}$ later
(See (\ref{splitness}) and below). Likewise, in the first two lines,
the remaining part $F'$ and $G'$ contain the information on the
leading order terms in $f$ and $g$: $S_i \cdot F'$ or $S_i \cdot G'$
respectively give the order of the polynomials of the dominant terms
in $f$ and $g$. If some polynomial is a complete square or complete
with some power, we say the singularity is split. Besides the gauge
part, the entire rest part of the canonical class, including that of
$E_8'$ takes part in the formation of the matter curves.

This decomposition shows only the information on dominant terms.
Knowing higher order terms means we have some information on an
enhanced group which embraces the current one. It follows that, for
a given group, vanishing of some parameter implies a symmetry
enhancement. We parameterize, for instance, one higher order than
(\ref{matbr}) in $f$ by the decomposition
\begin{equation}
 F = (\ord f_i + 1) S_i + (F' - S_i),
\end{equation}
 where we define a
new primed quantity for the new expansion as the last term in the
bracket. From the definition (\ref{Ddef}),
\begin{equation} \label{orderdelta}
 \ord \Delta \ge \min (3\,\ord f,2\,\ord g).
\end{equation}
So, we can know which of $f$ or $g$ is dominant in $\Delta$, hence
know whether $\Delta$ is to be increased. This explains the
inequality in Table \ref{t:fibclass}. The equality in
(\ref{orderdelta}) does not hold when the leading order in $\Delta$
is cancelled by those combinations of $f$ and $g$. For example, for
$SO(10)$ the enhanced group can be $SO(11)$ or $E_6$.\footnote{The
$SO(11)$ group is obtained from generic $D_6$ singularity without
splitness \cite{AG,Beetal}.} These correspond to the gauge symmetry
enhancement directions, when we send a certain term in $f$ or $g$ to
be zero. In such cases, information on an enhanced group is crucial.
For example, in $SO(10)$, some element of $\sum s_R D'_R$ comes from
a next dominant term $D_R'$ of $E_6$, indicating that the matter $R$
of $SO(10)$ is inherited from the branching of $R'$ of $E_6$ (See
Subsec. \ref{sec:so10}). In the next subsection, we will follow the
way of extensions.

In this work, we will consider a vanishing parameter that is {\em
not necessarily the leading} term, enhancing the symmetry in a
nontrivial way. There are also cases where the discriminant can be
factorized into two or more factors, each representing a simple
group or an Abelian group. So far, we have been interested in one
factor, or one simple group, which is enough for the local
unification group. We will make use of the fact that $D$ can show
not only the dominant contribution but also the the {\em full
components,} as the form (\ref{Ddecomp}) suggests. We will see that
the global factorization is easily catched by divisor expansion and
we can symmetrically describes all the factor group, including the
subgroups of the other $E_8$. It is not impossible to read off the
entire factors using the Tate's algorithm, which is mainly
interested in one factor \cite{Beetal,Ta}, if we go back to some
step and re-parameterize with respect to the position of another
factor, taking care of higher order terms in the discriminant.

\subsection{Weierstrass embedding and $E_8 \times E_8$}

For elliptic Calabi--Yau manifold $X_d$, usually we make further
{\em
  assumption} in the context of the duality to heterotic string:
$X_d$ allows a K3 fibration over a base $B_{d-2}$ which is
compatible with the elliptic fibration. Then, from the adiabatic
argument, it is fiberwise dual to the heterotic string on a
Calabi--Yau threefold, which is another elliptic fibration over the
same base $B_{d-2}$ \cite{Vafa:1996xn,Witten:1996bn}. Since this K3
fiber is elliptically fibered over $\P^1$, this requirement narrows
the possible base $B'_{d-1}$ of elliptic fibration:
\begin{equation} \label{hetdual}
 \text{$B'_{d-1}$ is a $\P^1$ fibration over $B_{d-2}$}.
\end{equation}
We will see that this requirement is quite strong, so that some
features, like the gauge symmetry breaking pattern, is less
sensitive to the choice of $B_{d-2}$. In fact such duality
essentially points that the unification group is that of the dual
heterotic string, namely $E_8 \times E_8$ or $SO(32)$, up to
possible enhancement from small instantons and dimensional
reduction. Relaxing this 
assumption to obtain authentic F-theory vacua is another
important question, which we shall not pursue here.\footnote{We have
a limited number of directly accessible Calabi--Yau fourfolds
\cite{Klemm:1996ts}. Recently uplifting of some class of type II
orientifold models to F-theory has been developed
\cite{Collinucci:2008zs,Blumenhagen:2009yv,Grimm:2009ef}. D-branes
and orientifolds are identified by divisors carrying an appropriate
symmetry, so that global consistency condition is derived.}

Mostly we will deal with compactification on Calabi--Yau threefold
$X_3$, $\pi': X_3 \to B_2'$ \cite{MV1,MV2,Bershadsky:1997zs,K3}.
Because of the requirement (\ref{hetdual}), the base manifold
can be completely specified to be the Hirzebruch surface $\F_n$, or
$\P^1$ bundle over $\P^1$. It is generated by two effective divisors
$r$ and $t$, satisfying the relations $r \cdot t = 1$, $r^2 = n$ and
$t^2 =0$.\footnote{Conventional notations: sometimes $\{C_0,C_\infty,f\}$ or
  $\{D_v,D_u,f\}$ for $\{r_0,r,t\}$ here.} Another irreducible divisor
    with negative 
self-intersection $r_0 = r - nt$ is disjoint from $r$
\begin{equation} \label{rnt}
 r \cdot r_0 = 0.
\end{equation}
The canonical class is
\begin{equation} \label{KFn}
 -K_{B_2'} = r + r_0 + 2t.
\end{equation}
We take $z'$ and $z$ as affine coordinates of the base $\P^1$ and
the fiber $\P^1$, respectively, such that $t=\{z'=0\}$ and $r =
\{z=0\}$. Near the base $\P^1 = r $, we have $-K_{B_2'}|_r=
-K_{B_2'} \cdot r = 2+n$. Since (\ref{rnt}) implies $r|_r = n$ and
$(r+nt)|_r = 0$, from the transformation of $f$ and $g$, we have
\begin{equation} \label{6dwei}
 y^2 = x^3 + x\sum_{j=-4}^4 f_{4\c_1-nj}(z')z^{4-j} + \sum_{k=-6}^6
 g_{6\c_1-nk}(z')z^{6-k},
\end{equation}
where $\c_1 \equiv \c_1(B_1 = \P^1)=2$. The coefficients
$f_{8-nj}(z')$ and $g_{12-nk}(z')$ are polynomials in $z'$ of degree
denoted by the subscripts, which we require to be non-negative,
otherwise we understand such terms do not exist. Immediately we see,
it is nothing but the deformation of $E_8 \times E_8$ singularity
\begin{equation} \label{e8wei}
 y^2 = x^3 + f_8(z') z^4 x + g_{12+n}(z') z^5 + g_{12} (z') z^6 +
 g_{12-n} (z') z^7.
\end{equation}
In (\ref{6dwei}), two pairs $z^{4\pm k}x$ and $z^{6 \pm k}$ are
respectively exchanged to each other with opposite sign, by $z \to
1/z$ up to rescaling $x \to z^4x $ and $y \to z^6 y$. This indicates
this affine form comes from that in the projective space, where we
can see the global structure \cite{MV2}. Thus $E_8 \times E_8$ is
the maximal symmetry we can see from the Weierstrass equation in
$z$. $f_8$ and $g_{12}$ are related to K\"ahler deformation on the
heterotic side, and we take both of them very large while fixing
$f_8^3/g_{12}^2$ finite \cite{MV2}. A quick check is that the possible
deformation has the dimension $8+12$ which is same as $h^{1,1}$ of
K3 in the heterotic dual.

In terms of divisors, two independent $E_8$ symmetries, or II$^*$
fibers, are supported by two disjoint curves $r$ and $r_0$, thanks
to the relation (\ref{rnt}). From Table \ref{t:fibclass}, each
II$^*$ singularity carries
 $\ord(f_i,g_i,\Delta_i)=(4,5,10)$, so as many $S_i$ are contained in $F,G,D$.
Then we have the following partitions
\begin{equation} \label{e8}
\begin{matrix}
 F =& -4K_{B_2'}  & = &  4r &+ & 4 r_0 &+& \underbrace{8t}_{F''}, \\
 G =& -6K_{B_2'}  & = &  5r &+ & 5 r_0 &+& \underbrace{r + r_0 + 12t}_{G''},\\
 D =& -12K_{B_2'} & = &  10 \underbrace{r}_{E_8} &+ & 10
 \underbrace{r_0}_{E_8'} &+&  \underbrace{2 r + 2r_0 + 24t}_{2D''_{\rm inst}}.
\end{matrix} \end{equation}
The last line shows the component of discriminant locus
(\ref{Ddecomp}). In view of $E_8$ supported by $r$, as in
(\ref{matbr}), the relations $r \cdot (4r_0+F'') = 8$ and $r \cdot
(5r_0+G'')=12+n$ shows the dominant terms in $f$ and $g$
respectively are $f_8 z^4x$ and $g_{12+n} z^5$. We have  similar
relations for the other $E_8'$, agreeing to (\ref{e8wei}). From the
discriminant locus, we see
\begin{equation}
 r \cdot (2r+2r_0 +24t) = 2(12+n), \quad
 r_0 \cdot (2r+2 r_0+24t) = 2(12-n),
\end{equation}
showing $g_{12 + n}^2 z^{10}$ and $g_{12-n}^2 z^{14}$ make up the
discriminant. Thus we identify the `instanton curve' as
\begin{equation}
r+r_0+12t.
\end{equation}
This is a divisor interacting with two different $E_8$s, which is
not easy to describe in the equation form. This connection leads to
the well-known consistency condition
\begin{equation} \label{Bianchi}
 n_1 + n_2 + n' = 24 \
 \Longleftrightarrow \ {\rm ch}_2(\V_1) + {\rm ch}_2(\V_2) +n'
 = \c_2 (\rm K3),
\end{equation}
where $n'$ is the number of possible blowups. 

At the zeros of $g_{12\pm n}(z')$, the $E_8$ singularity get worsen. This
is interpreted as the effect of small instantons in the heterotic side
\cite{Witten:1995gx,MV2}. To describe such broken symmetry, we need either
additional lower order terms of deformation, or blowing-up at that
point. For the former, being lower 
order terms, they do not affect the instanton terms not modifying the
relation (\ref{Bianchi}) staying at $n'=0$. 

We can blow-up in the
base to have zero size instantons. Blowing-up changes the
intersection number $n_1=12+n$ and $n_2=12-n$ to different values,
preserving the relation \cite{Bershadsky:1996nu}. In the low-energy
limit, the latter comprise the number of the tensor multiplets, not
counting the one containing the dilaton \cite{Seiberg:1996vs}. The
relation (\ref{Bianchi}) corresponds to Bianchi identity in the
heterotic side, where the instanton number is accounted by the
second Chern class of a vector bundle $\V_i$. It guarantees the
absence of anomalies. Each number of instantons $n_i$ and tensor
multiplets $n'$ from the blowup to be positive, so that each cannot
exceed 24. It is because the meaningful blowup is done at the
intersections between $E_8$ and the instanton curves. For $n'$, the
original $\F_n$ has no exceptional divisor of self-intersection
$(-1)$, by Castelnuovo--Enriques Criterion \cite{GH}, we cannot
blow-down the primitive $\F_n$. The negative instanton number can be
realized if we use analogous system as D7 and {\em anti}-D3 branes,
but we have to face stability issue. The only possible source of
larger symmetry than $E_8$ is using an exceptional divisors, from
the blow-up in the base, as components of the discriminant locus,
e.g. in Ref. \cite{AG,Bershadsky:1996nu}.

If we have two global sections instead one, we have $Spin(32)/\Z_2
\simeq SO(32)$ gauge symmetry. In this case, zero size instantons
give rise to gauge symmetry enhancement over the original group
\cite{AG,Witten:1995gx}. To obtain its subgroup is an interesting
topic.

\subsection{Four dimensions}

From the constraint of heterotic dual (\ref{hetdual}), the
Calabi--Yau fourfold has a similar structure as the above threefold
\cite{FMW}: This generalizes the construction of the Hirzebruch
surface. To describe the geometry globally, we use the language of
projective bundles. The base $B_3'$ of the elliptic fibration is the
total space of the projective bundle $\P(\O_{B_2} \oplus t)$, where
$t$ is a line bundle over $B_2$.\footnote{Again we use the same name
for the line bundle, its first Chern class and the associated
divisor.} Thus $B_3'$ is a $\P^1$ fibration over $B_2$, $\pi'': B_3'
\to B_2$. Take $r = 
\O(1)$ as the line bundle coming from the fiber $\P^1$. The sections
describing $r+t$ and $t$ have no common zeros, so that
\begin{equation} \label{rrt}
 r \cdot (r+t)=0.
\end{equation}

Of course, we can express the discriminant locus in the same way.
The total Chern class is, omitting pullback,
\begin{equation}
 \c(B_3') = \c(B_2) (1+r)(1+r+t)
\end{equation}
meaning that
\begin{align}
 \c_1(B_3') &= \c_1(B_2) + 2r + t, \label{1stchern}\\
 \c_2(B_3') &= \c_2(B_2) + \c_1(B_2)(2r+t) \label{2ndchern}.
\end{align}
Thus $ D = 12 \c_1(B_3') = 12 \c_1 (B_2) + 24r +12 t$. This is
analogous condition as (\ref{rnt}). This shows that, if we take the
base coordinate as $z$ such that $r=\{z = 0\}$, essentially the form
of Weierstrass equation, as a projective form, is the same as in six
dimensional case, with small replacement $f_{4 \cdot 2+nk} \to
f_{4\c_1(B_2)-kt}$ and $g_{6\cdot 2+nj} \to g_{6\c_1(B_2)-jt}$. Then the
Weierstrass equation essentially has the same form as (\ref{6dwei})
and the maximal gauge symmetry is $E_8 \times E_8$.

Knowing the details of the divisors supporting 7-branes, now
surfaces in the base $B_3'$ of Calabi--Yau fourfold, amounts to knowing
the form of $t$. Let us call the set of generators of the divisors
of $B_3'$, as $\{s_i\}$.
 We can replace $t$ with
the linear combinations of $s_i$'s
\begin{equation}
 t = \sum_{s_i \cdot r \ne 0} a_i s_i, \quad a_i > 0.
\end{equation}
The coefficients $a_i$ are inherited from the divisor relations
among $\pi''(r),\pi''(s_i)$ in $B_2$ \cite{Rajesh:1998ik}. For
example if we do a $\P^1$ fibration over $\F_n$, we have two more
parameters $k,m$ and $t=k \sigma + m f$ for the generators of the
divisors of $\F_n$. So a general divisor has a form $S_i = r +
a_\sigma \sigma + a_f f$ with some constants $a_\sigma,a_f$. Thus
$S_i$ have form $S_i = r + \sum k_i
s_i$  with $|k_i| < a_i$. We allow $k_i$ for negative integer,
interpreting as the intersection in the opposite orientation as in
six dimensional case.

In four dimensional compactification, the intersection of 7-branes
are surface in the base of elliptic fibration $B_3'$. To have four
dimensional fermions, we need more input giving rise to
four dimensional chirality, for example from the magnetic flux background
\cite{BHV,DW}. Also we have an extra condition on anomaly
cancellation on intersections having 3-brane charges, which is in
general singular \cite{Sethi:1996es,Becker:1996gj,Andreas:1999ng}.
Some components suffer worse singularity, whose remedy is recently
discussed in \cite{Collinucci:2008zs,Blumenhagen:2009yv}.

\section{Matter curves}

Once the decomposition of the discriminant locus (\ref{Ddecomp}) is
determined, like arranging 7-branes, we should be able to obtain the
matter content. From the mechanism of Katz and Vafa \cite{KV},
matter fields are localized along the intersections of two 7-branes.
For two gauge groups whose support are given by divisors $S_i$ and
$S_j$, we have the expansion
\begin{equation} \label{mattercurve}
 S_i \cdot S_j = \sum_{R,R'} w_{R_i,R'_j} \Sigma_{R_i,R'_j},
\end{equation}
where $\Sigma_{R_i,R'_j}$ are the effective intersections of the bases
spanning $S_i$ and $S_j$. The problem is, RHS of (\ref{mattercurve})
requires more information, that is, $w_{R,R'}$, while LHS gives at
best $\Sigma_{R,R'}$. In the next subsection, we will learn that the
relative ratio of $w_{R,R'}$ is determined by Green--Schwarz
mechanism. Then, in Subsec. \ref{sec:inher}, we consider this
problem considering enhanced gauge groups.

\subsection{Green--Schwarz mechanism}

The first guideline is the anomaly cancellation structure of
Green--Schwarz (GS) mechanism in six dimensions \cite{Sa}
\begin{align}
 \ell(\ad_i) - \sum_R \ell(R_i) n_{R_i} &= 6 K_{B'} \cdot
 S_i, \label{selfcond} \\
 y_{\ad_i} - \sum_R y_{R_i} n_{R_i} & = - 3 S_i^2, \\
 x_{\ad_i} - \sum_R x_{R_i} n_{R_i} &= 0 \label{irranom} \\
 \sum_{R,R'} \ell(R_i) \ell(R'_j) n_{R_i R'_j} &= S_i \cdot S_j. \label{relintersec}
\end{align}
Here $\ell(R)$ is the index of the representation $R$, ${\rm tr}_R
t_a t_b = \ell(R) \delta_{ab}$, and the quartic and the quadratic
invariants define the coefficients ${\rm tr}_R F^4 = x_R {\rm tr}
F^4 + y_R ({\rm tr} F^2)^2.$ By trace without subscript we mean the
minimal representation, i.e. the fundamental representation of
$SU(n)$. We use absolute normalization reflecting the branching
under the subgroup $
 R = \bigoplus (r,r') \ \Longrightarrow \ \ell(R) = \sum_r \ell(r)\dim r'
$
by defining property of trace. Thus, for the breaking $E_6 \to
SO(10)$, exactly $\ell({\bf 27})=\ell({\bf16})+\ell({\bf
10})+\ell({\bf 1})$ holds, for instance. Similar relations hold for
$x_R$ and $y_R$. Thus the condition (\ref{irranom}) gives us
well-known irreducible gauge anomaly cancellation. In Table
\ref{t:invs}, we displayed relevant invariants for some typical
gauge group and representations.

\begin{table}[l] \begin{center}
\begin{tabular}{|c|c|c|c|c|c|c|}
\hline
group & $R$ & multiplicity & $\ell(R)$ & $s_R$ & $x_R$ & $y_R$ \\
\hline
 $E_7$ & $\bf 133$ & 1 & 36 & $\cdot$ & 0 & 24\\\
 & $\frac12{\bf 56}$  & $8+n$ & $\frac12 12$ & 3 & 0 & $\frac12 6$ \\
\hline
$E_6$ & ${\bf 78}$  & 1 & 24 & $\cdot$ & 0 & 18  \\
 & ${\bf 27}$  & $6+n$ & 6 & 4 & 0 & 3 \\
\hline
 & ${\bf 45}$ & 1 & 16 & $\cdot$ & 4& 12 \\
$SO(10)$ & $\bf 16$ & $4+n$ & 4 & 3 & $-2$ & 3 \\
         & $\bf 10$ & $6+n$ & 2 & 2 & 2    & 0 \\
\hline
 & ${\bf N(2N-1)}$ & 1 & $4N-4$ & $\cdot$ & $4N-16$ & 12 \\
$SO(2N)$ & $\bf 2^{N-2}$ & $\cdot$ & $2^{N-3}$ & $\cdot$ & $-2^{N-4}$ & $3 \cdot 2^{N-5}$ \\
         & $\bf 2N$ & $\cdot$ & $N-1$ & $\cdot$ & 2    & 0 \\
\hline
 & $\bf 24$ & 1 & 10 & $\cdot$ & 10 & 6 \\
$SU(5)$ & $\bf 10$ & $2+n$ & 3 & 4 & $-3$ & 3 \\
 & $\bf 5$ & $16+3n$ & 1 & 1 &1 & 0 \\
\hline
  & $\bf N^2-1$ & 1 & $2N$ & $\cdot$ & $2N$ & 6 \\
$SU(N)$ & $\bf N(N-1)/2$ & $\cdot$ & $N-2$  & $\cdot$ & $N-8$ & $N-2$ \\
 & $\bf N$ & $\cdot$ & 1 & 1 &1 & 0 \\
\hline
\end{tabular}
\end{center}
\caption{Some group invariants, absolutely normalized so
  that the trace is the same for equivalent representations before and after
  branching. $s_R$ is the splitness of $E_8$ embedding.}
\label{t:invs}
\end{table}

In six dimensions, a matter curve is a point, so in our notation
(\ref{relintersec}) defines the expansion. We see that in the
decomposition of the matter curve the group factors are included.
For instance, $\bf 27$ of $E_6$ has $\ell({\bf 27})=6$, so in the
parallel separation $\bf (2,27)$ contributes $6 n_{\bf (2,27)}$ in
the decomposition. It is a generalization of the pair of mirror
branes on top of orientifold planes if we have vector representation
of $SO(n)$ type gauge symmetry. Eq. (\ref{selfcond}) shows the
matter content of a given brane is constrained by the product of the
cycle and the canonical class. These are already strong constraints
to limit possible models. In the simplest base without genus we have
no higher order representation than fundamental for $SU(n)$ case. On
the heterotic side, in the level one Kac--Moody algebra, we can have
spinorial and vector in $SO(2n)$ representation, and antisymmetric
tensor representations of $SU(n)$ \cite{Choi:2004vb}.

The GS conditions also give us information on what kind of divisor
can support the gauge group. They are derived \cite{Sa} from that a
number of antisymmetric tensor fields provide the Poincar\'e dual
basis to the first Chern class, entered in LHS of (\ref{Ddecomp})
and gauge groups are supported by the number of 7-branes, in RHS.
For $\F_n$ we have two two-cycles $h^{1,1}=2$, whose orthonormal
basis is provided by $r/\sqrt{n}$ and $r_0/\sqrt{n}$. We can
decompose
\begin{equation} \begin{split}
 \c_1 (B_2) &= \frac{n-2}{n} r + \frac{n+2}{n} r_0, \\
 S_i &= a_i\frac{r}{\sqrt n} + b_i \frac{r_0}{\sqrt n}.
\end{split}
\end{equation}
We can make 4-form which enters a factor in the anomaly polynomial
in $I_8$, $(n-2) \tr R^2/2\sqrt{n} +2 a_i \tr F_i^2, (n+2) \tr
R^2/2\sqrt{n} + 2 b_i \tr F_i^2$. Then we can cancel the anomaly by
counterterm including antisymmetric tensor fields $H^1 = {\rm
d}B^1+(-n+2)\omega_{3L}/2\sqrt{n} +   2a_i \omega_{3Y}^i, H^2 = {\rm
d}B^2 +(-n-2) \omega_{3L}/2\sqrt{n} + 2b_i \omega_{3Y}^i$, where
$\omega_{3L},\omega_{3Y}$ are respectively a gravitational and a
Yang--Mills Chern--Simons form. We note that, at least formally,
there is no requirement for $a_i,b_i$ to be positive integers or
rational number.

\subsection{Heterotic calculation and inheritance condition}
\label{sec:inher}

If a gauge group obtained from F-theory is semisimple, in general, it is not
sufficient to determine the multiplicity of a matter charged under
more than one simple group. It is because the relation
(\ref{relintersec}) gives only the total number of such matters for
given two groups.
 We can learn how to remedy 
it by studying the spectrum on the corresponding heterotic dual model.
The condition (\ref{hetdual}) implies that there exists a heterotic
dual for our construction on the Hirzebruch surface.
On the heterotic side, a semisimple group occurs if we have a line
bundle background. It has its own importance, since many realistic
models are obtained in this context.

First, we specify the embedding of the line bundle $\L$. The `Wilson
line' vector $V$ contains the information on the embedding in the Cartan
directions\footnote{For a similar construction in different context,
  see, e.g. \cite{Nibbelink:2007rd}.} 
\begin{equation}
 V= (V_1,V_2,\dots,V_8) \Longleftrightarrow \V = (\L^{V_1},\L^{V_2},\dots,\L^{V_8})
\end{equation}
for each $E_8$. 
Contrary to F-theory case where the gauge group is
arbitrary, here it is 
completely specified by breaking of $E_8 \times E_8$.
We require $\ch_2(\V) = \frac12 V^2 \c_1(\L)^2 = 12
\pm n$ for anomaly cancellation (\ref{Bianchi}). This data specifies
the spectral cover for $E_8$ \cite{FMW,Bershadsky:1997zs,DW}. 
For the standard weight vectors $w$ of $E_8$ \cite{Choi:2004vb}, the 
relation $V \cdot w =0$ determines the gauge bosons of unbroken group.

The matter spectrum is unambiguously computed by various bundle
cohomology groups. In our case, the only necessary information is the
charge $q=V \cdot w$ of the structure group.
From the branching ${\bf 248} \to
\bigoplus r_q$, the number of multiplet is obtained by index theorem,
\begin{equation} \label{hetindex} \begin{split}
 n_q &= \int_{\rm K3} {\rm Todd}({\cal M}) \ch (\V) \\
   &= \int \left( \frac{q^2}{V^2} \ch_2(\V) - \frac{1}{12}\c_2({\cal
   M})\right) \\
   &= \int \frac{q^2}{2} \c_1(\L)^2 - \frac1{12} \chi({\rm K3}) \\
   &= \frac{q^2}{V^2} (12 \pm n) - 2.
\end{split} \end{equation}
Physics is independent of the normalization,
because only the relative size $V/q$ matters which does not
change under Weyl reflections. We 
also used the Calabi--Yau condition $\c_1(\rm K3)=0$, and $\c_2(\rm
K3)=\chi(\rm K3)=24$ is the Euler number of K3. For the nonabelian
gauge bundle $\V$, we do not need the prefactor $q^2/V^2$ and
$\ch_2(\V)=\c_2(\V)$.

If there are more than two matter fields from the branching, there
is insufficient information for the relative relations on
$n_{R,R'}$. This happens because we have a {\em semisimple}
commutant to a $U(1)$ background. If we have a simple unbroken gauge
group, there is always a unique branching for a fixed $U(1)$ charge.
So we will use a principle, `{\em equal
inheritance condition,}' that the relative relation comes from the
unified group. Namely, if a 
given group is a common proper subgroup of two enhanced groups
$G_{M1} \cap G_{M2} = G$, a representation $R \in G$ is inherited by
branching of $R_{M1} \in G_{M1}$ and $R_{M2} \in G_{M2}$. Then, the
multiplicity is equally inherited from the original representations,
\begin{equation} \label{inhrepr}
 n_{R} = \frac12 n_{R_{M1}} + \frac12 n_{R_{M2}}.
\end{equation}
For example, $\bf (2,16)$ of $U(2)\times SO(10)$ comes from the
branching of ${\bf (2,27)}$ 
of $U(2) \times E_6$, and also from of ${\bf (3,16)}$ of $SO(10)
\times U(3)$. So we have the multiplicity $n_{\bf (2,16)}=\frac12
n_{\bf (2,27)} + \frac12 n_{\bf (3,16)}$. This is justified by the
index theorem 
\begin{equation}
 n_{q_1,q_2} = \int \left(\frac12 \frac{q_1^2}{V_1^2} + \frac12 \frac{q_2^2}{V_2^2} \right)(12+n) - 2.
\end{equation}
This happens only under the line bundle background, and the spectrum
is completely determined solely by the $U(1)$ charges. This
redefines the $U(1)$ normalization, under the condition that the
`instanton' number for the line bundle $(12+n)$ should not be
changed, and the second term on the RHS should always be the same
for the line bundle $- \c_2(\rm K3)/12$. In the limit $V_1=V_2$,
$q_1=q_2$ it should reduces to the original form (\ref{hetindex}). An
applied example is given at the end of Subsec. \ref{sec:so10}.

Finally, the absence
of six dimensional gravitational anomaly imposes the condition on
the number of tensor $n_T$, hyper $n_H$ and vector multiplets $n_V$,
\begin{equation} \label{gravanom}
 29 n_T + n_H - n_V = 273.
\end{equation}
It is automatically satisfied if the gauge symmetry is obtained from
breaking of $E_8$, satisfying the constraint, LHS of
(\ref{Bianchi}). The dimension of the moduli space of $E_8$ with the
instanton number $n_1 = 12+n$ is $\dim {\cal M}_{E_8}(n_1) =
30n_1-248=30n+112$. It does not change by spontaneous symmetry
breaking, so that $n_H - n_V$ is preserved. A blow-up increases the
number of tensor multiplet so that $n_T = h^{1,1} - 1$, and
$n_T=n'+1$ for $\F_n$. With 20 deformations from $f_8$ and $g_{12}$,
 we have the relation $n_H - n_V = 20 + (n_T - 1) + \dim{\cal
M}_{E_8}(n_1) + \dim{\cal M}_{E_8}(n_2) = 273 - 29n_T$ automatically
satisfying the constraint (\ref{gravanom}).

\subsection{$u(1)$ brane}

We always have to face a $U(1)$ symmetry if we consider a semisimple
group. The easiest way to understand this is using the heterotic
dual. A background line bundle yields unbroken group as the
commutant in the unified group. This $U(1)$ commutes to itself and
survives \cite{Nibbelink:2007rd,Blumenhagen:2005pm}.

In the perturbative limit, the $U(1)$ group is related to the
overall center of momentum motion of a D-brane stack. Thus there is
no independent degree of freedom for this. Since we obtain such
symmetry if we make a subleading parameter to be zero, we may also
employ a virtual 7-brane $Q_j$ responsible for the $U(1)$ symmetry
as the corresponding divisor for the parameter set to be zero. We
have a modified relation corresponding to (\ref{relintersec})
\begin{equation} \label{u1intersec}
\sum_{R,q} \ell(R_i) \frac{q_j^2}{V^2}  n_{R_i q_j} = S_i \cdot Q_j.
\end{equation}
It follows that the very existence of such brane predicts that the
RHS should be universal for every other group.

\subsection{Brane reduction}

From the eight dimensional twisted supersymmetry, we have the field
$\phi_{mn}$ transforming as $K_{B'} \otimes \ad P$ on $B'$
\cite{BHV}. Its nontrivial profile $\langle \phi_{mn} \rangle$ is
naturally interpreted as the deformation of the 7-branes in the
normal direction. We do not need to know its field theoretic
description in detail, since we have explicit description of gauge
group in terms of Weierstrass equation. The point is that
$\phi_{mn}$ transforms as the adjoint representation. So we expect
at least for simplest profile we have gauge symmetry breaking of
$E_8$ without rank reduction. We have analogy for the D-brane stack
in the flat space, where the constant VEVs of normal scalar to the
D-brane correspond to parallel separation of the branes. Higgsing it
leads us to the symmetry breaking and enhancement $G \leftrightarrow
H_1 \times H_2$, the adjoint branches as
\begin{equation}
 \adj G \leftrightarrow (\adj H_1,{\bf 1}) + ({\bf 1}, \adj H_2) + \text{(off-diagonal comp.) + c.c. }
\end{equation}
The matter representations come from the branching of the adjoint.
The well-known example is $U(m+n) \to U(m) \times U(n)$.
$$
 \bf (m+n)^2-1 \to (m^2-1,1) \oplus (1,n^2-1) \oplus (1,1) \oplus
 (m,n) \oplus (\overline m, \overline n).
$$
The off-diagonal elements $\bf (m,n)$ or $\bf (\overline m,\overline
n)$ are in general not vectorlike because each belong to different
cohomology with different dimension. It depends on how the
transition takes place. If the symmetry breaking is spontaneous, we
can think of the reverse process as symmetry enhancement. The
general form is
\begin{equation} \label{Amndef}
 y^2 = x^2 + \prod_{l=1}^m(z - u_l) \prod_{k=1}^n(z-t_k).
\end{equation}
This is viewed as a local form of the Weierstrass equation. If all
$u_l$ and $t_k$ assume the same value, say 0, the symmetry is
enhanced to $A_{m+n-1}$
\begin{equation} \label{Amn}
 y^2 = x^2 + z^{m+n},
\end{equation}
i.e., $\ord(f,g,\Delta)=(0,0,m+n)$. Thus $u_l$ and $t_k$ in
(\ref{Amndef}) parameterize deformations of (\ref{Amn}). Among
deformations, if all $u_l$ have the same value $0$, and all $t_k$
have another same value $t \ne 0$, the deformed curve describes
$A_{m-1} \oplus A_{n-1}$ symmetry
\begin{equation} \label{AmAn}
 y^2 = x^2 + z^m (z-t)^n.
\end{equation}
Around $z = 0$, $(z-t)$ is fixed to be nonzero, thus the curve looks
as that of $A_{m-1}$. The same holds for $A_{n-1}$ around $z = t$.
In particular if $t$ is a complex number, the surfaces $\{z=0\}$ and
$\{z=t\}$ is homologous. We can generalize is to the case of
exceptional groups \cite{KM,Katz:1996xe}. In terms of divisors, we
have
\begin{equation}
 (m+n)S  \longrightarrow m S_1 + n S_2
\end{equation}
The divisors $S,S_1,S_2$ may not be same or linearly equivalent.

There are some components which are not responsible for the gauge
dynamics. The deformation due to finite size instanton is rather
close to brane recombination, a la \cite{BHV}, triggered by the
field on the defect developing VEV. In D-brane case, $k$ instantons
embedded in $U(N)$ is described by $k$ D0 on $N$ D4 branes (or its
T-duals) \cite{BVS,Un}. If we just place D0s, we just have zero size
instantons and there is gauge symmetry enhancement to $U(k)\times
U(N)$. If we grow the size of instantons, it is translated into
assigning nonzero VEVs to $\bf (k,N)$ representations. The
size-growing of instanton is clearer if we take T-dual in some two
direction, then the initial system looks like two stacks of
intersecting branes and the bifundamental fermion $\bf (k,N)$ is one
localized at the angles. The resulting gauge symmetry is broken down
to $U(k-l)\times U(N-l)$ undergoing the rank reduction. We describe
this case
$$
 k S_1 + N S_2 \longrightarrow (k-l) S_1 + (N-l) S_2 + l (S_1+ S_2)
$$
where the last term does not support the gauge degree of freedom.
The corresponding Weierstrass equation is obtained
\begin{equation}
 y^2 = x^2 + z^k z^{\prime N} \to y^2= x^2 + z^{k-l} z^{\prime N-l} \prod_{i=1}^l (z z'+b_i).
\end{equation}
Here we set deformation parameter with poles of order $l$
\cite{BHV}, because coordinate dependent deformation on the
off-diagonal components can be diagonalized with poles. Needless to
say, $b_i \to 0$ recovers the original symmetry. This is not trivial
if we embed this curve in $E_8$, where we will see that we have a
very large parameter forcing $b_i \to 0$. See Subsec.
\ref{sec:e6a1}.

The splitness $s_R$ contains information on, to which group the
given group is embedded. If the gauge group at hand is embedded in
exceptional group, it is far from trivial since the notion of
parallel is not trivial in general. Consider the above transition
\begin{equation} \label{kNinitial}
 D = k S_1 + N S_2 + D''.
\end{equation}
Since $k S_1 + D''$ and $NS_2 + D''$ are I$_0$ singularities, we
have $n_{\bf k} = S_1 \cdot (N S_2 + D'') / s_{\bf k}$ and $n_{\bf
N} = S_2 \cdot (k S_1 + D'') / s_{\bf N}$. In the next section (and
also in Ref \cite{Beetal}) we see $s_{\bf k}$ is not always 1. Since
we know $n_{\bf k,N} = S_1 \cdot S_2$, we have
\begin{equation} \begin{split}
 &s_{\bf k} n_{\bf (k,1)} = N (s_{\bf k}-1) S_1 \cdot S_2  + S_1 \cdot
 D'', \\
 &s_{\bf N} n_{\bf (1,N)} = k (s_{\bf N}-1) S_1 \cdot S_2  + S_1 \cdot
 D''. \end{split}
\end{equation}
The recombination is done by giving VEVs to $l$ multiplets of
$(k,N)$. We have the branching ${\bf (k,1)} \to {\bf (k-l,1)} +
l{\bf (1,1)},\ {\bf (k,N)} \to {\bf (k-l,N-l)} + l{\bf (1,N-l)}
+l{\bf (k-l,1)}+l^2{\bf (1,1)}$. Thus we know
\begin{equation} \label{nkminusl}
 n_{\bf k-l} = n_{\bf (k,1)} + N n_{\bf (k,N)}
\end{equation}
and a similar for $n_{\bf N-l}$. The divisor (\ref{kNinitial})
undergoes the reduction to
\begin{equation}
 D =  (k-l)S_1 + (N-l)S_2 + \underbrace{D'' + l (S_1 + S_2)}_{D''_{\rm new}},
\end{equation}
where $D''_{\rm new}$ makes up a new divisor. Thus, we have
\begin{equation}
 s_{\bf k-l} n_{\bf k-l} = S_1 \cdot ((N-l) S_2 + D''_{\rm new})
 =  N S_1 \cdot S_2 + S_1 \cdot D'' + l S_1^2 .
\end{equation}
Equating it with (\ref{nkminusl}), we have
$$ l S_1^2 = N(2-s_{\bf k-l} - s_{\bf k}) n_{\bf (k,N)}
 + (s_{\bf k}-s_{\bf k-l}) n_{(\bf k,1)}.  $$
If we assume $s_{\bf k}=s_{\bf k-l} =1$, which is the case in the
perturbative $SU(n)$ type group, immediately we see the
self-intersection of $S_1$ should be zero. This is the case when we
embed D-branes on flat torus where parallel branes have no
intersection. It also means that we cannot embed these $U(k)$ groups
on the divisor $r$ of Hirzebruch space $\F_n$ with $n \ne 0$.

We may generalize such mechanism for exceptional group. For the
reduction of the divisor, the orders of polynomials $\ord
(f,g,\Delta)$ are preserved. If the Dynkin diagram allows, there is
always corresponding symmetry breaking \cite{Choi:2003pqa}. The
remaining information is monodromy or splitness conditions. To see
collision between different gauge groups, we further decompose the
discriminant locus.
\begin{equation} \label{moredecomp}
 D =  - 12 K_{B'} =  \sum_i (\ord \Delta_i) S_i + \sum_{k,U(1)} Q_k +
\dots,
\end{equation}
where, inevitably we always have a 7-brane $Q_k$ parameterizing
$U(1)$ dynamics. Still, in the description employing the divisors
and intersections, we can symmetrically discuss the divisors
supporting factor groups.

We have two independent ways of obtaining the matter spectrum. In
particular if there is only one kind of matter representation $R$,
the product $n_R = S_i \cdot D''/s_i = (\ell(\ad_i)-6 K_{B'} \cdot
S_i)/\ell(R)$. In this way we can inversely reconstruct Tate's
algorithm, since knowing the redundancy $s_R$ for each matter $R$
completely specifies the required splitness. Plugging the GS
conditions (\ref{selfcond})-(\ref{relintersec}) into (\ref{matbr}),
\begin{equation} \label{splitness}
   6 \ell(\ad_i) -  \ord \Delta_i y_{\ad_i} = \sum_R \big( 6
   \ell(R_i) - \ord \Delta_i y_{R_i} - 3 s_{R_i} \big)
  n_{R_i}.
\end{equation}
Since we know all the information from the group theory and the
orders of singularity, we can obtain the splitness from this
equation. If there are several branched representations we go to
enhanced group to input more information.

\section{Extended gauge symmetry} \label{sec:extsymm}

In this section, we first review the engineering of simple subgroups
of $E_8$ on the Hirzebruch surface \cite{Beetal}, reinterpreted in
terms of intersection theory \cite{K3}. From the discussion in the
previous section, we can identify the gauge symmetry by the data
(\ref{matbr}). We can generalize the description to semisimple, 
Abelian and non-simply laced group.

\subsection{$E_6$}
The first example is $E_6$ and we can similarly analyze $E_7$. We
assume that the corresponding singularity is located along the
divisor $r$. Later we will relate the mother group $E_8$ in the
previous section, sitting along the same divisor. From Table.
\ref{t:fibclass}, $E_6$ has $\ord(f,g,\Delta)=(3,4,8)$. Thus we may
write down two partitions
\begin{equation} \label{e6}
\begin{matrix}
 F = & 3 r &+& \underbrace{r + 4r_0  +  8t}_{F'} , \\
 G = & 4 r &+& \underbrace{2r + 6r_0 +  12t}_{G'},\\
 D = & 8\underbrace{r}_{E_6} &+& \underbrace{4r+12r_0+24t}_{4D'_{\bf 27}} .
\end{matrix}
\end{equation}
From $r \cdot F' =  n+8$, the dominant term in $f$ is
$f_{8+n}(z')z^3x$, and from $r \cdot G' = 2n+12$ we see the dominant term
in $g$ is $g_{12+2n}(z')z^4$, which is also the leading term in the
discriminant. It is seen from $r \cdot 4D'_{\bf 27} = 4 n + 24 = r
\cdot 2 G' = 2(2n+12)$. The splitting condition tells us that the
matter curve is further reduced to $(2r+12t)/2 = r+6t$, or
$g_{12+2n}=q_{6+n}^2$. The resulting equation is
\begin{equation} \label{e6wei}
 y^2 = x^3 + f_{8+n} z^3 x + q_{6+n}^2 z^4 + O(z^5).
\end{equation}
The discriminant has the form
\begin{equation}
 \Delta = 27 z^8 q_{6+n}^4(z') + O(z^9).
\end{equation}
We have $(6+n)$ localized matters $\bf 27$ along the zeros of
$q_{6+n}(z')$. Heterotic string on K3 independently gives such the
result, for which higher order terms in $O(z^9)$ are irrelevant,
i.e. if we embed an instanton in $SU(3)$, we have the same spectrum.
However this is just {\em one} of the possible solutions, only when
we assume that the higher order terms are {\em generic}, i.e. not
factorized any more. We will consider more general case in the
following.

We can blow-up at the intersection points. Like instanton, we
do not need details of the position, but the embedding group and the
total number of blow-up sufficiently specify the physics.
Suppose we blow-up at $n'$ points, then the relation changes the
relations $\hat r^2 = n-n', \hat t^2 = -n', -  \hat K \cdot \hat r =
-n - 2+n'$ while leaves $\hat r \cdot \hat t =1$. Thus the number of
$\bf 27$ becomes $\hat r \cdot (\hat r + 6 \hat t) = n-n'+ 6$.
We see that still there is no anomaly.

\subsection{$E_6 \times U(2)$ and factorization} \label{sec:e6a1}

In (\ref{e6wei}), note that
tending $q_{6+n} \to 0$ enhances the symmetry to $E_7$,
\begin{equation*}
 y^2 = x^3 + f_{8+n} z^3 x  + O(z^5).
\end{equation*}
If we had no $f_{8+n} z^3 x$ term in (\ref{e6wei}), the gauge
symmetry would be $U(2) \times E_6$, because then $q_{6+n} \to 0$
enhances the symmetry to $E_8$. We will see later we also need
$g_{12+n} = 0$, clearly interpreted as zero instanton number in the
heterotic side.  Considering full factors up to $O(z^6)$ in the
equation, it is described by
\begin{equation} \label{su3e6}
 y^2 = x^3 + f_8 z^4 x + q_{6+n}^2 z^4 + g_{12} z^6 + O(z^7).
\end{equation}
Keeping to $O(z^{10})$ the discrimiant becomes\footnote{In terms of
\cite{KM}, the deformation
  parameters satisfy the relation
  $f_{E_8}(U;t)=f_{E_6} \cdot f_{A_1}(U-\tau_1/3-\mu_1/9;t'')$.}
\begin{equation} \label{disce6su2}
 \Delta = 54 z^8 q_{6+n}^2 g_{12} \big(z^2 + q_{6+n}^2/2g_{12} \big) + O(z^{11}).
\end{equation}
The factor in the bracket shows the degree two $A_1$ singularity
factor $z^2$ up to finely broken effect by $q_{6+n}^2/g_{12}$, since
we are working in the limit $g_{12} \to \infty$. If we really remove
$q_{6+n}$ and restore $g_{12+n}$ term, we again recover $E_8$. Later
we will see evidences for surviving $SU(2)$ with explicit examples.
This unattractive form is due to the fact that we have deformation
in $y$, see for example Ref. \cite{KV}. This is formally
parameterized as $\ord(f,g)=(0,0)$ for I$_n$ or $A_{n-1}$
singularity, and to see the full symmetry, we should refer to the
equation in Tate's form, which we do not need for our discussion
\cite{Beetal,Ta}.

We restate the above in terms of divisors. Since we have no
$f_{8+n}z^3 x$ term, the corresponding divisor $F$ has is expanded
around $4r$, instead of $3r$. Still $\ord \Delta = 8$ as in
(\ref{orderdelta}). Putting $2r$ for $A_1$ singularity, the
remaining part is $2r+24t+2r_0$. Since we have no terms except
$f_{6+n},f_8$ and $g_{12}$, only we can expand as
\begin{equation} \label{e6a1}
\begin{matrix}
 F = & 4 r &+& 0 &+& 4r_0 & +  & 8t , \\
 G = & 4 r &+& 0 &+& 5r_0 &+& 2(r +6t)+ r_0,\\
 D = & 8\underbrace{r}_{E_6} &+& 2\underbrace{r}_{A_1}
   &+& 10 \underbrace{r_0}_{E_8'}
 &+& 2(r+6t) + 2r_0 +12t .
\end{matrix} \end{equation}
We assume the other $E_8'$ is unbroken for convenience, which is in
fact unnecessary since all the divisors for $E_6$ and $SU(2)$ is
orthogonal to $r_0$ supporting $E_8'$. Therefore we can interpret
this situation as the symmetry breaking of $E_8$ via transition
\begin{equation}
 10 r \longrightarrow 8 r + 2 r.
\end{equation}
Two $r$'s on the RHS respectively support $E_6$ and $SU(2)$, which
are not necessarily the same curves, as (\ref{disce6su2}) shows;
Better say, they are {\em different} curves that are {\em linearly
equivalent.} On $\F_n$, linearly equivalent curves can have nonzero
net intersections. Their intersection number is $n$, so we have as
many $\bf (2,27)$ divided by the group theoretical factor $\ell({\bf
2})\ell({\bf 27})=6$, as in (\ref{relintersec}). Obviously, VEVs of
two $\bf (2,1)$'s would completely break $SU(2)$, reducing to the
previous $E_6$ model having $(n+6)$ ${\bf 27}$'s, {\em neglecting}
the other factor group $SU(2)$. Thus we have net number of $\bf
(1,27)$ as $(n+6)-\frac{n}{6} (\dim {\bf 2})$. Similarly we can
calculate the number of $\bf (2,1)$'s $16+6n-\frac{n}{6} (\dim {\bf
27})$, therefore
\begin{equation} \label{e6a1spec}
 \textstyle \frac{n}{6}{\bf (2,27)}_1 + \frac13(18+2n){\bf
   (1,27)}_{-2} + \frac12(32+3n){\bf (2,1)}_{-3}.
\end{equation}
We verify that it is consistent with the GS relations
(\ref{selfcond})-(\ref{irranom}). Sometimes fractional multiplicity
indicates localization of the matter along a part of the geometry.
In our calculation the divisor is globally given, so we can have
meaningful spectrum if $n$ is a multiple of 6. Only the positive
matter multiplicity is allowed in six dimension, since there is only
one possible chirality for the matter. When $n = 0$ we have no
localized $\bf (2,27)$, which is analogous to `parallel separation'
of D7-branes in the perturbative description without bifundamental
zero mode.

We singled out $u(1)$ divisor $(r+8t)$ in (\ref{e6a1}), provided by
the missing $f_{8+n}$ term in (\ref{u1intersec}). We can track its
origin from the enhanced group $E_7 \to E_6 \times U(1)$. As a
consistency check, we calculate the GS conditions with respect to
$E_6 \times U(1)$ and $SU(2) \times U(1)$ respectively giving
\begin{equation} \begin{split} \textstyle
 \frac{n}{6}(\dim {\bf 2}) \ell({\bf 27}) \cdot 1^2 &+ \textstyle
 \frac13(18+2n)(\dim{\bf 1}) \ell({\bf 27})\cdot (-2)^2 =
 18(n+8), \\ \textstyle
 \frac{n}{6}(\dim {\bf 27}) \ell({\bf 2}) \cdot 1^2 &+ \textstyle
 \frac{1}{2}(32+3n)(\dim {\bf 1}) \ell({\bf 2})\cdot (-3)^2
 = 18(n+8).
\end{split} \end{equation}
From the difference
of the number of charged hypermultiplets and vector multiplets, we
have
\begin{equation} \begin{split}
 [n_{H} - n_V]_{E_6\times U(2)} &= \textstyle
 \frac{n}{6}\cdot 2\cdot 27 + \frac 13(18+2n) \cdot 27 + \frac12
 (32+3n) \cdot 2 - 78 - 3 -1 \\
 &= 30n + 112,
\end{split} \end{equation}
showing that these are the symmetry obtained from $E_8$. Since there
is no singlet hypermultiplets, we have consumed all the moduli; We
have to tune all the coefficients to have the unbroken group. This
also indicates that $E_6 \times U(2)$ is the maximal subgroup
including $E_6$.\footnote{According to Ref. \cite{KM}, this is
allowed as the maximal, for the purely algebraic reason.}
Note on the fixed moduli: $q_{6+n}$ is
completely tuned, leaving no massless field.

On the heterotic side, this model corresponds to one under the line
bundle background. There is no zero in $g_{12+n}(z')=0$, meaning
that the second Chern class is $\c_2=0$ for the background bundle.
In this context we have $U(2) \times E_6$ as the maximal subgroup
containing $E_6$.
 To provide the appropriate amount to the Bianchi
identity (\ref{Bianchi}), we interpret that the line bundle $\L$
gives $\ch_2(\V)= V^2 \c_2(\L)^2/2 = 12+n$, with $V^2=(2,1,1,0^5)^2
= 6$. Every subgroup here commutes to $U(1)$ in $E_8$ to which the
line bundle is embedded. Using the index theorem (\ref{hetindex}),
we can calculate the spectrum on the heterotic side
$$ n_q = q^2 (12+n)/V^2 - 2, $$
thus
\begin{equation*} \begin{split}
 ({\bf 2,27})_1   &: n_1 = \textstyle \frac16 (12+n) - 2 = \frac{n}{6},\\
 ({\bf 1,27})_{-2}&: n_{-2} = \textstyle \frac{(-2)^2}{6} (12+n) - 2 = \frac13(18+2n),\\
 ({\bf 2,1})_{-3} &: n_{-3} = \textstyle \frac{(-3)^2}{6}(12+n)-2 = \frac12
 (32+3n),
\end{split} \end{equation*}
 agreeing with
(\ref{e6a1spec}). It is not a coincidence that
the matter multiplicity and charge quantization is very similar. We
can understand the line bundle plays a similar role as the
instanton, giving the dimension of the moduli space $30 \ch_2(\V)
-248 = 30n +112$.


This description shows that we can investigate the behavior of both
singularities of $E_6$ and $A_1$ on the equal footing. The role of
divisors carrying each of them is equal. In this example, two groups
are supported by linearly equivalent divisors, so except the order
of singularity the matter multiplicity was symmetric. There is a
case where the subgroup of $E_8$ is not supported along the original
divisor for $E_8$.

\subsection{Deformation of positions and matching the full $E_8 \times E_8$}

It is not compulsory for $E_6$ to lie along the divisor $r$. The GS
condition showed that the divisor supporting a gauge field is just
expanded by two-cycles with arbitrary coefficients. We may consider
for example $(r-2t)$,
\begin{equation} \label{e6r2t}
\begin{matrix}
 F = & 3 (r-2t) &+& \underbrace{r + 4r_0  +  14t}_{F'} , \\
 G = & 4 (r-2t) &+& \underbrace{2r + 6r_0 +  20t}_{G'},\\
 D = & 8\underbrace{(r-2t)}_{E_6} &+& \underbrace{4r+12r_0+40t}_{4D'_{\bf 27}} .
\end{matrix} \end{equation}
From the intersection number $(r-2t) \cdot (4r+12r_0+40t) = 4(n+2)$
with the same splitness 4. Thus the number of ${\bf 27}$ is $n+2$.
The spectrum is consistent with
GS conditions (\ref{selfcond})-(\ref{relintersec}).
We know that, from the $E_7$ mother group, we have order $n+6$
polynomial for $r+8t$. From the $E_8$ mother group, we have instanton
number $(r-2t) \cdot (r+r_0+12t) = 8+n$. We can verify the total
dimension of the moduli space
$$ (n+2) \cdot 27 - 78 + (8+n)+(4+n)+ (2+n)  = 30n-8. $$

This dimension is also obtainable from a model with $E_6 \times U(2)$
all localized at $(r-2t)$. The spectrum is
\begin{equation}
 \textstyle \frac16(-4+n){\bf (2,27)}_1+\frac13(10+2n){\bf
   (1,27)}_{-2}+\frac12(20+3n){\bf (2,1)}_{-3}.
\end{equation}
The total moduli space has the dimension $30n-8$.

The Weierstrass equation can see only the partial information. On
$z=0$, we can see only the projected part for the divisor $(r-2t)|_r
= (r-2t)\cdot r = n-2$. This means that we cannot count the right
degree of freedom. For example the instanton number is not
$(r+r_0+12t)|_r = 12+n$, as shown just before. Since the instanton
number is $8+n$, it seems that we can reproduce the same spectrum by
redefining the instanton number $8+n \equiv 12+ n'$. Comparing the
spectrum in the previous subsection, indeed we reproduce the
spectrum. Our present model should be viewed as one originating
$E_8$ located at $(r-2t)$, whose moduli space is $30n'+112=30n-8$.
The physics should be equivalent since it is completely specified by
the instanton embedding. However, the relative relation between two
$E_8$s are different. For example, putting the other $E_8'$ on
$r_0$, we have relative intersection number $(r -2t) \cdot r_0 = -2$
signaling the inconsistency. Also The gravitational anomaly
cancellation seems difficult, however we can show always we can. 

If we place the other $E_6' \times U(2)'$, as a subgroup of $E_8'$ on,
$(r_0 + 2t)$, we have a similar spectrum
\begin{equation}
 \textstyle \frac16(4-n){\bf (2,27)}_{1}+\frac13(26-2n){\bf
   (1,27)}_{-2}+\frac12(44-3n){\bf (2,1)}_{-3}.
\end{equation}
so the total moduli space is again $(30n-8)+(-30n+232)=224$, again
leading to gravitational anomaly cancellation. This condition can be
tracked back that two $E_8$ subgroups should be independent
$$(r-2t) \cdot (r_0+2t) = 0.$$
Although formally anomalies cancel, it is consistent if all the
coefficients are nonnegative. Since the coefficients of $\bf (2,27)$
are of opposite signs, so only $n=4$ case seems valid. As we noted,
this is the `parallel separation' condition. It is also possible a
different subgroup of $E_8'$ can give the desired dimension of the
moduli space $-30n+232$. If we put all the hidden sector gauge group
along the divisor $(r_0 +2t)$ this condition is valid. Since the
total dimension of the moduli space does not change by spontaneous
symmetry breaking, we can construct the subgroup in the hidden
sector. Again, this shows that the gravitational anomaly
cancellation comes from the embedding to $E_8$.

We can consider more general embedding of $E_6\times U(2)$ singularity
in a form
\begin{equation}
 D = 8(r+at) + 2 (r+bt) + \dots.
\end{equation}
In case $a \ne b$, there is no easy argument that the model comes
from $E_8$, since there can be nonzero intersection with a divisor
supporting the hidden sector group, i.e. subgroup of the other
$E_8'$. In other words, it is hard to construct the mirror model,
where all the visible brane is disjoint from the hidden brane.
Usually at least one of the intersections has a negative
intersection number, and even if positive, it should be proportional
to the number of charged matter under both groups, weighted by group
theoretical factors. If the group is not small enough, the total sum
of the coefficients exceeds 24, which is required by the Calabi--Yau
condition $D=-12K_{B'}$. However for such small group, the origin
can also be tracked from the subgroup of $SO(32)$. With our method,
it seems not possible to construct the model which is not the
subgroup of $E_8 \times E_8$. At best one $E_8$ can carry the sum of
divisors $12r +2rt$. We can state such condition as
\begin{equation}
  12 r + 24t - \sum_{i \in E_8 \text{ subgroup}} (\ord \Delta_i ) S_i
 \ge 0
\end{equation}
where the inequality means the divisor on LHS is effective.
Also, if the total gauge symmetry lies outside $E_8$ unified group,
we have no physical reason for anomaly cancellation
(\ref{gravanom}).

\subsection{$SO(10)$ and its extended groups} \label{sec:so10}
It suffices to study $SO(10)$ for more general case. The splitness
condition implies the relation \cite{Beetal}
\begin{equation} \label{so10wei}
g_{12+3n} = 2 s_{4+n}^3, \quad f_{8+2n} = - 3 s_{4+n}^2, \quad
\textstyle g_{12+2n} = q_{6+n}^2 - f_{8+n} s_{4+n},
\end{equation}
so that
\begin{equation} \label{so10disc}
 \Delta = 108 z^7 s_{4+n}^3 q_{6+n}^2 + O (z^8).
\end{equation}
As before for generic $f_{8+n}$ we have only $SO(10)$, however for
$f_{8+n}\to 0$ we have gauge symmetry  
enhancement $SO(10)\times U(2)\times U(1)$ without changing the
leading order form (\ref{so10disc}). We have rank one symmetry
enhancement by $s_{4+n} \to 0$ to $E_6$. Instead if we send $q_{6+n}
\to 0$, the singularity is generic $D_6$ without splitting, which
describes $SO(11)$ gauge symmetry. Further splitness condition
imposing inter-relations among $g_{12+n},f_{8+n},s_{4+n}$ yields
$SO(12)$.

The pure $E_6$ theory cannot be a unification group, since then $\bf
10$ is not possible from the branching of the adjoint $\bf 78$. Also
from the form of equation, we have $SO(11)$ enhancement direction,
implying a larger symmetry. The structure of Weierstrass equation
requires the $E_8$ embedding.

In terms of divisors of $\F_n$, we have
\begin{equation} \label{so10}
\begin{matrix}
 F = & 2 r &+& 2r + 4r_0  +  8t, \\
 G = & 3 r &+& 3r + 6r_0 +  12t,\\
 D = & 7\underbrace{r}_{D_5} &+& \underbrace{5r+12r_0+24t}_{D'} .
\end{matrix} \end{equation}
From the products $r \cdot (2r + 4 r_0 + 8t) = 2(n+4)$ and $r \cdot
(3r+6r_0+12t)=3(n+4)$ we draw out the above leading order relations
of $g_{12+3n},f_{8+2n}$ have a special splitting condition. Since
they cancel out each other in the discriminant, we rely on the next
leading order terms in $z$, which is that of $E_6$ in the previous section.
Thus we inherit the same number of  $\bf 10$ of $SO(10)$ from the
branching of $\bf 27$ of $E_6$, whereas two $\bf 16$'s of $SO(10)$
is absorbed by Higgs mechanism. $D'$ is decomposed into
$2(r+6t)+3(r+4t)$, up to irrelevant $r_0$'s. With respect to $r$,
they have respectively $(n+6)$ and $(n+4)$ intersections, we have
\begin{equation} \label{so10spec}
 (n+6) {\bf 10} + (n+4) {\bf 16}
\end{equation}
localized along the corresponding intersections.

In the same way, we can calculate $SO(10) \times U(3)$
spectrum. From the branching from $E_8$, we have no $\bf (10,1)$, so
the only vector multiple under the $SO(10)$ is $\bf (10,3)$ whose
number is then $(n+6)/3$. We have
\begin{equation} \textstyle
 \frac{1}{12}(n-12){\bf (16,3)}_{-1}
   + \frac13(n+6){\bf(10,3)}_{-2}
   + \frac13(4n+42){\bf (1,3)}_{-4}
   + \frac14(3n+28)({\bf 16,1})_3.
\end{equation}
We can check that the matter curve is again $(r+8t)$, since we set
$f_{8+n}=0$ to have the desired symmetry enhancement. Indeed we
verify (\ref{u1intersec})
$$ \sum_{R,q} \frac{q^2}{V^2} \ell(R) n_{R,q} = 3(n+8), $$
which is universal with respect $SO(10) \times U(1)$ and $SU(3)
\times U(1)$. In the heterotic side we use the vector $V^2 =
(2,2,2,0^5)^2=12$, so that the matter spectrum is also obtained by
the index (\ref{hetindex}) $n_q =  q^2 (12+n) /12 - 2$.

Finally, consider a semisimple group $SO(10) \times U(2) \times
U(1)$. This is the common subgroup of $E_6 \times U(2)$ and $SO(10)
\times U(3)$. These enhanced groups are very useful in understanding
the structure of the subgroup. In terms of divisors,
\begin{equation} \label{e6u2u1p}
 10 r  \longrightarrow \left \{ \begin{matrix} 8 r + 2r \\ 7r + 3r \end{matrix}
 \right \} \longrightarrow 7 r + r + 2r.
\end{equation}
From the branching of the adjoint of $E_8$, we have more than one
matter charged under both $SO(10)$ and $SU(2)$. We use the
inheritence condition (\ref{inhrepr}). Considering the enchanced
groups $E_6 \times SO(10)$ and $SO(10) \times U(3)$ we have, for
example, ${\bf (2,16)}_{1}$ comes from the branching of ${\bf
(2,27)}_{1}$ of $E_6 \times U(2)$, and also from of ${\bf
(3,16)}_{-1}$ of $SO(10) \times U(3)$. So we have the multiplicity
$n_{(\bf 2,16)}=\frac12 n_{\bf (2,27)} + \frac12 n_{\bf (3,16)}$.
Therefore we have highly nontrivial spectrum
\begin{align*}
 & \textstyle \frac{1}{8}(n-4){\bf (2,16)}_{1,1}+
   \frac14(n+4){\bf (2,10)}_{1,2}+
   \frac{1}{12}(n-12){\bf (2,1)}_{1,0}\\
 +& \textstyle \frac{1}{24}(17n+156){\bf
   (1,16)}_{-2,3}+\frac12(n+8){\bf
   (1,10)}_{-2,-2}
 +(10+n) {\bf(1,1)}_{-2,4} \\
 +&\textstyle \frac{1}{12}(17n+180) {\bf
   (2,1)}_{3,4}+\frac{1}{24}(n-36){\bf (1,16)}_{0,1}.
\end{align*}
At this stage, the spectrum seems not be realistic since $n$ should be
a multiple of 12 which is also larger than 36. A certain rearrangement of
the divisor supporting the group, or blowing-up is necessary. In what
follows, we are content with formal check of consistency
conditions. The dimension of the moduli space is $30n+112$.
It satisfies GS condition
$$ \ell({\bf 2})\ell({\bf 16}) n_{\bf (2,16)} + \ell({\bf 2})
\ell({\bf 10}) n_{\bf (2,10)} = n, $$
and consistent with the Higgsing that leaving only $SO(10)$
\begin{align*}
 n_{\bf 16} &= 2 n_{\bf (2,16)} + n_{{\bf (1,16)}_{-2,3}}+n_{{\bf (1,16)}_{0,1}}
= n+4, \\
 n_{\bf 10} &= 2 n_{\bf (2,10)} + n_{\bf (1,10)} = n+6.
\end{align*}
We can check the matter curve relation (\ref{u1intersec}) is
universal for any of four combinations between $SO(10),SU(2)$ and
$U(1),U(1)'$.

This group is obtained from the above $SO(10)$ by making $f_{8+n} =
0, g_{12+n} = 0$. The resulting discriminant is
\begin{equation}
 \Delta = z^7 \Big[ (-36f^2_8 s^2_{4+n} z^3
   + 54 \big(g_{12} q^2_{6+n} z^2 + 2 g_{12} s^3_{4+n} z + 2 f_8 s^4_{4+n} ) \big)z
  + 27 q^4_{6+n} z + 4 s^3_{4+n} q_{6+n}^2 \Big]
\end{equation}
In terms of $f_8 \to \infty$ and $g_{12} \to \infty$ with
$f_8^3/g_{12}^2$ fixed, we find a hierarchy in orders $O(f^2_8)$ and
$O(f_8)\sim O(g_{12})$, therefore we have the factorization $\Delta
\sim z^7 \cdot z^2 \cdot z$.

\subsection{$F_4$}
We describe a non-simply laced group $F_4$, which is described by
{\em generic} $E_6$. So it carries the same orders as $E_6$,
\begin{equation} \label{f4}
\begin{matrix}
 F = & 3r &+&     \underbrace{r+4r_0 + 8t}_{F'} \\
 G = & 4r &+& \underbrace{2r + 6r_0 + 12t}_{G'},\\
 D = & 8\underbrace{r}_{F_4 = E_6} &+&
 \underbrace{4r+10r_0+20t}_{2 D'_{\bf 26}}+\underbrace{2r_0+4t}_{D'_{\bf 1}} .
\end{matrix} \end{equation}
The non-splitness condition tells us that $D'$ is not proportional
to $G'$, so it should be decomposed. As always there is a
unimportant ambiguity in the distribution of $r_0$. We have only
splitness 2 from Tate's algorithm. The product of $F_4$ divisor and
$D'$ shows we have $2 \cdot (2n+10)$ intersections.

It seems not possible to deform the Weierstrass equation into the
desirable form, because the Weierstrass equation shows the
singularity form when the {\em entire} the subgroups of $E_8$ is
lying on the original position $r$. We note that the next order term
plays a role. We see $f_4$ originated from the tuning $f_4 \cdot
g_{12+2n} = f_{8+n}^2$ agreeing with $D_1' = (2F' - G')$ and $r
\cdot D_1' = 4$. Therefore we have $(2n+10)/2=(n+5)$ matter
representations in $\bf 26$ and four $\bf 1$s. A similar thing
happens for a {\em generic} $A_1$ singularity in Ref. \cite{Beetal},
where we have `antisymmetric' representation $\bf 1$ of $SU(2)$ with
multiplicity $(4+2n)/2=n+2$. Even if we have difficulty in
expressing a generic group with monodromy reduction, we can
calculate the multiplicity from the intersection theory.

\section{Conclusion}

We have illustrated how to specify a gauge theory and obtain its
matter spectrum in F-theory.
The essential problem is how to decompose the
discriminant locus, in 
(\ref{matbr}) and (\ref{moredecomp}). It is analogous to configuring 
D7-branes in the internal manifold. 
Duality to heterotic string
(\ref{hetdual}) suggests that our manifold should be compatible to both
elliptic and K3 fibration.
This limits the possible gauge groups as
ones in heterotic string, $E_8 \times E_8$ or $SO(32)$, from 
independent considerations of Weierstrass equation or the conservation of
7-brane charges. Most of
vacua admits an interpretation that they are obtained by symmetry
breaking of the unified group. Also, several constraints of model
building, such as anomaly cancellation conditions from
Green--Schwarz mechanism, the formation of matter branes and the
dimension of moduli space, indicate the unification towards the
above unification group. This unification condition is only evaded by
blowing-up in the base of 
elliptic fibration and including the resulting exceptional divisors in the
discriminant locus. Even in this case, the above unification group
is a good starting point to consider top-down approach.
Usually by top-down
approach, we meet many unwanted charged matters, as well as hidden
sectors used for symmetry breaking.

A matter curve comes from the intersection between 7-branes.
If we have explicit information on 7-branes, not one for local
unification group but the complete set of 7-branes in the theory,
we can obtain the matter curve without ambiguity.
To see this we considered semisimple gauge group.
In Section \ref{sec:extsymm}, we took examples having
semisimple gauge group. In every case, spectrum and moduli space
matches perfectly to that of $E_8 \times E_8$ in both
F-theory (geometric) and heterotic (gauge bundle) side. In the
latter, we can calculate the spectrum
using index theorem, whose vacuum is parameterized by spectral cover. 

The requirements from Green--Schwarz mechanism
(\ref{selfcond})-(\ref{relintersec}) shows that, in the expansion
of the intersection of two 7-branes, each matter curve (as a divisor in
7-brane support) is weighted by
a group theoretical factor depending on the localized matter. 
The nontriviality comes if we embed the groups in the exceptional
group, since we have no notion of parallel separation of exceptional
branes, unlike that of D-branes. Thus we cannot determine the 
spectrum solely by geometric data of the intersection.
We obtained the following rule, from the corresponding
heterotic dual model in Subsec. \ref{sec:inher};
To completely specify the matter spectrum, we should consider every
possible enhanced gauge groups; the matter multiplicity is equally inherited
from those enhanced group. Thus knowing the global structure is again
important.

We also have 7-branes responsible for
$U(1)$, which do not directly give the matter multiplicity, as in
(\ref{u1intersec}). The anomaly constraint confirms the existence of
such $U(1)$ brane. Such $U(1)$'s provided additional constraints on
the matter coupling (implicitly used in \cite{Mo}).

In this way we can find many nontrivial vacua of F-theory with
semisimple group, opening up more possibility for model building.
They admit heterotic duals with 
line bundle backgrounds, some of which are close to many models
suggested so far. An additional group 
outside the conventional unification group of $SU(5)$ or $SO(10)$ can
also play a role, either providing one source of the interaction of
the Standard Model in the larger unification group, or supplementing
additional constraints.

\subsection*{Acknowledgements}
The author is grateful to Ralph Blumenhagen, Teruhiko
Kawano and Tatsuo Kobayashi, and particulary to Tae-Won Ha
for useful discussion. He is supported in part
by the Grant-in-Aid for Scientific
Research No. 20$\cdot$08326 and 20540266 from the Ministry of
Education, Culture, Sports, Science and Technology of Japan.

\end{document}